\documentclass[pre, aps, showpacs, showkeys]{revtex4}
\usepackage{graphicx}
\begin{document}
%\draft

\title{Nonlinear theory of dust lattice mode coupling in
dust crystals \footnote{Proceedings of the \textit{International
Conference on Plasma Physics - ICPP 2004}, Nice (France), 25 - 29
Oct. 2004; contribution P2-062; available online at:
\texttt{http://hal.ccsd.cnrs.fr/ccsd-00001893/en/} .}}

\author{I. Kourakis\footnote{On leave from: U.L.B. - Universit\'e Libre
de Bruxelles, Physique Statistique et Plasmas C. P. 231, Boulevard
du Triomphe, B-1050 Brussels, Belgium; also: Facult\'e des
Sciences Apliqu\'ees - C.P. 165/81 Physique G\'en\'erale, Avenue
F. D. Roosevelt 49, B-1050 Brussels, Belgium;
\\Electronic address: \texttt{ioannis@tp4.rub.de}} and P. K.
Shukla\footnote{Electronic address: \texttt{ps@tp4.rub.de}}}
\affiliation{Institut f\"ur Theoretische Physik IV,
Fakult\"at f\"ur Physik und Astronomie, \\
Ruhr--Universit\"at Bochum, D-44780 Bochum, Germany}

\date{\today}

\begin{abstract}
Quasi-crystals formed by charged mesoscopic dust grains (dust
lattices), observed since hardly a decade ago, are an exciting
paradigm of a nonlinear chain. In laboratory discharge
experiments, these quasi-lattices are formed spontaneously in the
sheath region near a negative electrode, usually at a levitated
horizontal equilibrium configuration where gravity is balanced by
an electric field. It is long known (and experimentally confirmed)
that dust-lattices support linear oscillations, in the
longitudinal (acoustic mode) as well as in the transverse, in
plane (acoustic-) or off-plane (optic-like mode) directions.
Either due to the (typically Yukawa type) electrostatic
inter-grain interaction forces or to the (intrinsically nonlinear)
sheath environment, nonlinearity is expected to play an important
role in the dynamics of these lattices. Furthermore, the coupling
between the different modes may induce coupled nonlinear modes.
Despite this evidence, the elucidation of the nonlinear mechanisms
governing dust crystals is in a rather preliminary stage. In this
study, we derive a set of (coupled) discrete equations of motion
for longitudinal and transverse (out-of-plane) motion in a one
dimensional model chain of charged dust grains. In a continuum
approximation, i.e. assuming a variation scale which is larger
than the lattice constant, one obtains a set of coupled modified
Boussinesq-like equations. Different nonlinear solutions of the
coupled system are discussed, based on localized travelling wave
ans\"{a}tze and on coupled equations for the envelopes of
co-propagating quasi-linear waves.
\end{abstract}

\pacs{52.27.Lw, 52.35.Fp, 52.25.Vy}

\keywords{Dusty (Complex) Plasmas, Dust Crystals, Lattice Modes,
Soft Condensed Matter.}

\maketitle

\section{Introduction}

Recent studies of various collective processes in dust
contaminated plasmas (DP) \cite{PSbook} have been of significant
interest in relation with linear and nonlinear waves which are
observed in laboratory and space plasmas. An issue of particular
importance is the formation of strongly coupled DP crystals by
highly charged dust grains, for instance in the sheath region
above a horizontal negatively biased electrode in experiments
\cite{PSbook, Morfill}. Low-frequency oscillations may occur in
these mesoscopic dust grain quasi-lattices, in both longitudinal
(acoustic mode) \cite{Melandso} and transverse (in-plane shear
acoustic mode, off-plane optic-like mode) directions, as
theoretically predicted and experimentally observed (see in Ref.
\cite{PSbook} for a review).

In this paper, we focus on the nonlinear description of dust grain
displacements in a one-dimensional dust crystal, which is
suspended in a levitated horizontal equilibrium position where
gravity and electric (or, possibly magnetic \cite{Yaro}) forces
balance each other. Considering the coupling between the
horizontal ($\sim \hat x$) and vertical (off-plane, $\sim \hat z$)
degrees of freedom, and an arbitrary inter-grain interaction
potential form $U(r)$ (e.g. Debye or else) and sheath potential
$\Phi(z)$ (not necessary parabolic), we aim in deriving a set of
equations which should serve as a basis for forthcoming studies of
the nonlinear behaviour of longitudinal and transverse dust
lattice waves (LDLWs, TDLWs) propagating in these crystals. The
relation to recent studies of a similar scope (here recovered as
special cases) is also discussed.

\section{The model}

Let us consider a layer of charged dust grains (mass $M$ and
charge $q$, both assumed constant for simplicity) of lattice
constant $r_0$. The Hamiltonian of such a chain is of the form
\[H = \sum_n \frac{1}{2} \, M \, \biggl( \frac{d \mathbf{r}_n}{dt}
\biggr)^2 \, + \, \sum_{m \ne n} U(r_{nm}) \, +
\Phi_{ext}(\mathbf{r}_n)\, ,
\]
where $\mathbf{r}_n$ is the position vector of the $n-$th grain;
$U_{nm}(r_{nm}) \equiv q \, \phi(x)$ is a binary interaction
potential function related to the electrostatic potential
$\phi(x)$ around the $m-$th grain, and $r_{nm} =
|\mathbf{r}_{n}-\mathbf{r}_{m}|$ is the distance between the
$n-$th and $m-$th grains. The external potential
$\Phi_{ext}(\mathbf{r})$ accounts for the external force fields in
which the crystal is embedded; in specific, $\Phi_{ext}$ takes
into account the forces acting on the grains (and balancing each
other at equilibrium, ensuring stability) in the vertical
direction (i.e. gravity, electric and/or magnetic forces); it may
also include the parabolic horizontal confinement potential
imposed in experiments for stability \cite{Samsonov} as well as,
for completeness, the initial laser excitation triggering the
oscillations in experiments.

 %figure X
\begin{figure}[htb]
 \centering
\resizebox{9cm}{!}{ \includegraphics[]{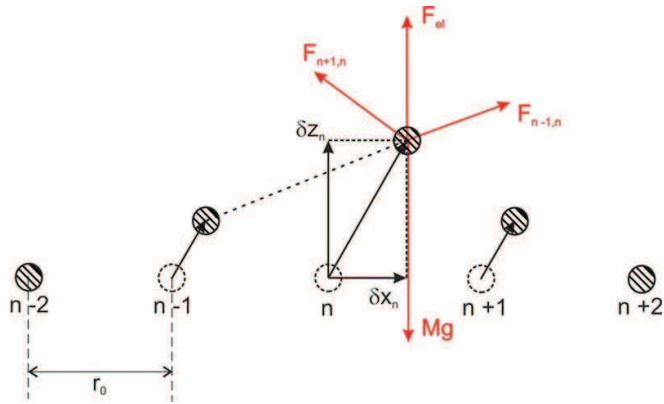} }
\caption{\small Dust grain vibrations in the longitudinal ($\sim \hat x$)
and transverse ($\sim \hat z$) directions, in a 1d dust lattice.} \label{SS1}
\end{figure}

\subsection{2d equation of motion}

Considering the motion of the $n-$th dust grain in both the {\em
longitudinal} (horizontal, $\sim \hat x $) and the {\em
transverse} (vertical, off--plane, $\sim \hat z$) directions (i.e.
suppressing the transverse in-plane -- shear -- component, $\sim
\hat x$), so that $\mathbf{r}_n = (x_n, z_n)$, we have the
two-dimensional (in $x, z$) equation of motion
\begin{equation}
M\, \biggl( \frac{d^2 \mathbf{r}_n}{dt^2} \, + \nu \, \frac{d
\mathbf{r}_n}{dt} \biggr) = \,- \sum_n \,\frac{\partial
U_{nm}(r_{nm})}{\partial \mathbf{r}_n} \, + \,
\mathbf{F}_{ext}(\mathbf{r}_n) \, \equiv \, q \,
\mathbf{E}(\mathbf{r}_n) \, + \, \mathbf{F}_{ext}(\mathbf{r}_n) \,
, \label{eqmotion0}
\end{equation}
where $E_j(x) = - \partial \phi(\mathbf{r})/\partial x_j$ is the
(interaction) electrostatic field and $F_{ext, j} = - \partial
\Phi_{ext}(x)/\partial x_j$ accounts for all external forces in
the $j-$ direction ($j = 1/2$ for $x_j = x/z$); the usual
\textit{ad hoc} damping term was introduced in the left-hand-side
of Eq. (\ref{eqmotion0}), involving the damping rate $\nu$ due to
dust--neutral collisions.

\subsection{Nonlinear vertical confining potential}

We shall assume a smooth, continuous variation of the (generally
inhomogeneous) field intensities $\mathbf{E}$ and/or $\mathbf{B}$,
as well as the grain charge $q$ (which may vary due to charging
processes) near the equilibrium position $z_0 = 0$. Thus, we may
develop
\[
E(z) \approx E_0 \, + \, E'_0 \, z \,+ \, \frac{1}{2} E''_0 \, z^2
\, + \, ... \, ,
\]
\[
B(z) \approx B_0 \, + \, B'_0 \, z \,+ \, \frac{1}{2} B''_0 \, z^2
\, + \, ... \, ,
\]
and
\[
q(z) \approx q_0 \, + \, q'_0 \, z \,+ \, \frac{1}{2} q''_0 \, z^2
\, + \, ... \, ,
\]
where the prime denotes differentiation with respect to $z$ and
the subscript `$0$' denotes evaluation at $z = z_0$, viz. $E_0 =
E(z=z_0)$, $E'_0 = d E(z)/d z|_{z=z_0}$ and so forth. Accordingly,
the electric force $F_e = q(z) E(z)$ and the magnetic force $F_m =
- \partial (m B)/\partial z = - 2 \alpha \, B \, \partial
B/\partial z$ (where the grain magnetic moment $\mu$ is related to
the grain radius $a$ and permeability $\mu$ via $m = (\mu - 1) a^3
\,B/(\mu + 2) \equiv \alpha B$ \cite{Jackson}), which are now
expressed as
\[
F_e(z) \approx q_0 E_0 \, + \, (q_0 E'_0 + q'_0 E_0) \, z
% \qquad \qquad \] \[ \qquad \qquad
\,+ \, \frac{1}{2} (q_0 E''_0 + 2 q'_0 E'_0 + q''_0 E_0) \, z^2 \,
+ \, ... \, ,
\]
and
\[
F_m(z) \approx - 2 \alpha B_0 B'_0 \, - \, 2 \alpha ({B'_0}^2 +
B_0 {B''}_0)\, z \,
%\qquad \qquad \] \[ \qquad \qquad
- \, \alpha (B_0 B'''_0 + 3 B_0' {B''}_0) \, z^2 \, + \, ... \, ,
\]
may be combined to give
\[
F_e + F_m = - \frac{\partial \Phi}{\partial z} \, ,
\]
where we have introduced the phenomenological potential $\Phi(z)$
\begin{eqnarray}
\Phi(z) & \approx & \Phi(z_0) \, + \frac{\partial \Phi}{\partial
z}\biggr|_{z=z_0} \, z + \, \frac{1}{2!}\, \frac{\partial^2
\Phi}{\partial z^2}\biggr|_{z=z_0}\, z^2
%\nonumber \\ & & \qquad \qquad
+ \, \frac{1}{3!}\, \frac{\partial^3 \Phi}{\partial
z^3}\biggr|_{z=z_0}\, z^3 + \, ...
\, \nonumber \\
& \equiv & \Phi_0 \, + \Phi_{(1)} \, z + \, \frac{1}{2}\,
\Phi_{(2)} \, z^2 + \, \frac{1}{6}\, \Phi_{(3)} \, z^3 + \, \cdots
\, . \label{Phi}
\end{eqnarray}
The definitions of $\Phi_{(j)}\equiv \bigl( {\partial^j
\Phi(z)}/{\partial z^j}\bigr|_{z=z_0} \,= \, - (q E_0)^{(j-1)}_0 +
\alpha (B^2)^{(j)}_0 $ (here, the superscript within parenthesis
obviously denotes the order in partial differentiation; $j = 1, 2,
...$) are obvious:
\begin{eqnarray}
\Phi_{(1)} & = & - (q E)_0 + \alpha (B^2)'_0 \, = - q_0 E_0 +
2 \alpha B_0 B'_0
\nonumber \\
\Phi_{(2)} & = & - (q E_0)'_0 + \alpha (B^2)''_0 \,
\nonumber \\
& = & - (q'_0 E_0 + q_0 E'_0) +
2 \alpha ({B'}_0^2 + B_0 B''_0)
\nonumber \\
\Phi_{(3)} & = & - (q E_0)''_0 + \alpha (B^2)'''_0 \,\nonumber \\
& = & - (q''_0 E_0 + 2 q'_0 E'_0+ q_0 E''_0) +
2 \alpha (3 B'_0 B''_0 + B_0 B'''_0) \, ,\nonumber \\
& & \label{defPhij}
\end{eqnarray}
and so forth. Obviously, $\Phi_{ext} = \Phi - M g z$. The
(vertical) force balance equation $\partial \Phi_{ext}/\partial z
= 0$, viz.
\[
M g = q_0 E_0 - 2 \alpha B_0 B'_0 \, ,
\]
is satisfied at equilibrium.

\section{Discrete equations of motion}

Assuming small displacements from equilibrium, one may Taylor
expand the interaction potential energy $U(r)$ around the
equilibrium inter-grain distance $l r_0 = |n-m| r_0$ (between
$l-$th order neighbors, $l=1, 2, ...$), i.e. around $\delta x_n
\approx 0$ and $\delta z_n \approx 0$, viz.
\[ U(r_{nm}) = \sum_{l'=0}^\infty \,\frac{1}{l'!}
\biggl. \frac{d^{l'} U(r)}{d r^{l'}} \biggr|_{r = l \, |n-m| r_0}
\, (x_{n} - x_{m})^{l'} \, ,\] where $l'$ denotes the degree of
nonlinearity involved in its contribution: $l' = 1$ is the linear
interaction term, $l' = 2$ stands for the quadratic potential
nonlinearity, and so forth. Notice that the inter-grain distance
$r = [(x_n - x_m)^2 + (z_n - z_m)^2]^{1/2}$ also needs to be
expanded, i.e. near $|x_n - x_m| = l r_0$ and $z_n - z_m = 0$, so
that $\partial U(r)/\partial x_j = (\partial U(r)/\partial r)
(\partial r/\partial x_j) \approx ...$. Obviously, $\delta x_n =
x_n - x_n^{(0)}$ and $\delta z_n = z_n - z_n^{(0)}$ denotes the
displacement of the $n-$th grain from the equilibrium position
($x_n^{(0)}, z_n^{(0)}) = (n r_0, \, 0)$. Retaining only
nearest-neighbor interactions ($l = 1$), we obtain the coupled
equations of motion
\begin{eqnarray}
\frac{d^2 (\delta x_n)}{dt^2} \, + \nu \, \frac{d (\delta
x_n)}{dt} = \, \omega_{0, L}^2 \, (\delta x_{n+1} + \delta
x_{n-1} - 2 \delta x_{n}) \qquad \qquad \qquad \qquad \qquad
\qquad
\nonumber \\
\, - a_{20}
\, \biggl[ (\delta x_{n+1} - \delta x_{n})^{2} - (\delta x_{n} -
\delta x_{n-1})^{2} \biggr]
\nonumber \\
\, + \, a_{30} \, \biggl[ (\delta
x_{n+1} - \delta x_{n})^{3} - (\delta x_{n} - \delta x_{n-1})^{3}
\biggr]
%\nonumber \\
+ \, a_{02}
\, \biggl[ (\delta z_{n+1} - \delta z_{n})^{2} - (\delta z_{n} -
\delta z_{n-1})^{2} \biggr]
\nonumber \\
\, - a_{12}\, \biggl[ (\delta
x_{n+1} - \delta x_{n}) (\delta z_{n+1} - \delta z_{n})^{2} -
(\delta x_{n} - \delta x_{n-1}) (\delta z_{n} - \delta
z_{n-1})^{2} \biggr]
\, ,\nonumber \\
\label{discrete-eqmotion-x}
\end{eqnarray}
and
\begin{eqnarray}
\frac{d^2 (\delta z_n)}{dt^2} \, + \nu \, \frac{d (\delta
z_n)}{dt} = \, \omega_{0, T}^2 \, (2 \delta z_{n} - \delta z_{n+1}
+ \delta z_{n-1}) \, - \,\omega_g^2 \, \delta z_{n} \qquad \qquad
\qquad \qquad \qquad \qquad
\nonumber \\
- \, K_1 \,
(\delta z_{n})^2 - \, K_2 \, (\delta z_{n})^3 \, +
\frac{a_{02}}{r_0} \, \biggl[ (\delta z_{n+1} - \delta z_{n})^{3}
- (\delta z_{n} - \delta z_{n-1})^{3} \biggr]
\nonumber \\
\, + \, 2 \, a_{02}\, \biggl[ (\delta
x_{n+1} - \delta x_{n}) (\delta z_{n+1} - \delta z_{n}) - (\delta
x_{n} - \delta x_{n-1}) (\delta z_{n} - \delta z_{n-1}) \biggr] \,
\nonumber \\
- \, a_{12}\, \biggl[ (\delta x_{n+1} - \delta x_{n})^2 (\delta
z_{n+1} - \delta z_{n}) - (\delta x_{n} - \delta x_{n-1})^2
(\delta z_{n} - \delta z_{n-1}) \biggr] \, ,
\label{discrete-eqmotion-z}
\end{eqnarray}
where we have defined the longitudinal/transverse oscillation
characteristic frequencies
\begin{equation}
\omega_{0, L}^2 = \, U''(r_0)/M \, , \qquad \omega_{0, T}^2 = -
\,U'(r_0)/(M r_0) \, , \label{frequencies}
\end{equation}
(both assumed to be positive for any given form of interaction
potential $U$) and the quantities
\begin{eqnarray}
a_{20} = - \frac{1}{2 M}\,U'''(r_0) \, , \qquad a_{02} = -
\frac{1}{2 M r_0^2}\, \bigl[ U'(r_0) - r_0 U''(r_0) \bigr] \, ,
\nonumber \\
a_{30} = \frac{1}{6 M}\,U''''(r_0) \, , \qquad a_{12} = -
\frac{1}{M r_0^3}\, \bigl[ U'(r_0) - r_0 U''(r_0) + r_0^2 \,
\frac{1}{2} \,U'''(r_0) \bigr] \, , \label{coefficients}
\end{eqnarray}
which are related to coupling nonlinearities. The {\textit{gap
frequency}} $\omega_g$ and the nonlinearity coefficients $K_1$ and
$K_2$ are related to the form of the sheath environment (i.e. the
potential $\Phi$) via
\begin{equation}
\omega_g^2 = \Phi_{(2)}/M \, , \qquad K_1 = \Phi_{(3)}/(2 M) \, ,
\qquad K_2 = \Phi_{(4)}/(6 M) \, . \label{K12-def}
\end{equation}
Obviously, the prime denotes differentiation, viz. $U''(r_0) =
\biggl. {d^2 U(r)}/{d r^2} \bigr|_{r = r_0}$ and so on. In the
above equations of motion, we have distinguished the linear
contributions of the first neighbors from the nonlinear ones, i.e.
the first line in the right--hand--side from the remaining ones,
in both equations. Note that all of the coefficients are defined
in such a way that they bear {\em{positive}} values for
Debye--type interactions, i.e. if $U_D(r) = (q^2/r) \exp(-
r/\lambda_D)$ ($\lambda_D$ is the effective Debye length) since
odd/even derivatives are then negative/positive; however, the sign
of these coefficients is not a priori prescribed for a different
interaction potential $U(r)$. Indeed, we insist on expressing all
formulae in such a manner that a different interaction law may
easily be assumed in a \textit{``plug--in''} manner; in
particular, even though the Debye potential $U_D$ is widely
accepted in DP crystal models, we think of the modification of $U$
when one takes into account a magnetic field \cite{Yaro} or the
ion flow towards the negative electrode \cite{Ignatov}.
Nevertheless, we provide the explicit form of the coefficients
$a_{ij}$ defined above for a Debye potential, for clarity, in the
Appendix.

Upon careful inspection of the discrete equations of motion above,
one notices that the lowest order nonlinearity in the longitudinal
motion is due to the intergrain interaction law, while
nonlinearity in the vertical motion is primarily induced by the
coupling to the horizontal component (and, to less extent, by
interactions).

\section{Continuum approximation}

Adopting the standard continuum approximation, we may assume that
only small displacement variations occur between neighboring
sites, i.e.
\[
\delta x_{n \pm 1} \approx u \pm r_0 \frac{\partial u}{\partial
x} + \frac{1}{2} r_0^2 \frac{\partial^2 u}{\partial x^2} \pm
\frac{1}{3!} r_0^3 \frac{\partial^3 u}{\partial x^3} +
\frac{1}{4!} r_0^4 \frac{\partial^4 u}{\partial x^4} \pm \, ... ,
\]
where the (horizontal) displacement $\delta x_n (t)$ is now
expressed via a continuous function $u = u(x, t)$. The analogous
continuous function $w = w(x, t)$ is defined for $\delta z_n (t)$.

One may now proceed by inserting this ansatz in the discrete
equations of motion (\ref{discrete-eqmotion-x},
\ref{discrete-eqmotion-z}), and carefully evaluating the
contribution of each term. The calculation, quite tedious yet
perfectly straightforward, leads to a set of coupled continuum
equations of motion in the form
\begin{eqnarray}
\ddot{u} \,+ \, \nu \, \dot{u} - c_L^2 \, u_{xx} -
\frac{c_{L}^2}{12}\, r_0^2 \, u_{xxxx} \, = \,- \, 2 \, a_{20}\,
r_0^3 \, u_x \,u_{xx} \, + \, 2 \, a_{02}\, r_0^3 \, w_x \,w_{xx}
\,
\nonumber \\
- \, a_{12}\, r_0^4 \, [(w_x)^2 \,u_{xx} + 2 w_x w_{xx} u_x ] \,
+ \, 3 \, a_{30}\, r_0^4 \, (u_x)^2 \,u_{xx} \, ,
\label{eqmotion-gen-continuum-x}
\end{eqnarray}
\begin{eqnarray}
\ddot{w} \,+ \, \nu \, \dot{w} + c_T^2 \, w_{xx}\, +
\frac{c_{T}^2}{12}\, r_0^2 \, w_{xxxx}\,+ \, \omega_g^2 \, w \, =
\, - \, K_1 \, w^2 \, - \, K_2 \, w^3 \,
\nonumber \\
+ \, 2 \, a_{02}\, r_0^3 \,( u_x \,w_{xx} \, + \,w_x \,u_{xx} )
\nonumber \\
+ \, 3 \, a_{02}\, r_0^3 \, (w_x)^2 \,w_{xx} -\, a_{12}\, r_0^4 \,
[ (u_x)^2 \,w_{xx} + 2 u_x u_{xx} w_x ] \, ,
\label{eqmotion-gen-continuum-z}
\end{eqnarray}
where higher-order nonlinear terms were omitted. We have defined
the characteristic velocities $c_L = \omega_{0, L}\, r_0$ and $c_T
= \omega_{0, T}\, r_0$; the subscript $x$ denotes partial
differentiation, so that $u_x \, u_{xx} = (u_x^2)_x/2$ and
$(u_x)^2 \, u_{xx} = (u_x^3)_x/3$. Remember that the {\textit{gap
frequency}} $\omega_g$ and the coefficients $K_1$ and $K_2$ are
related to the form of the sheath electric and/or magnetic
potential via (\ref{K12-def}) above, viz. $F_{el} = M \, g - M \,
\omega_g^2 \, z - K_1 \,z^2 - K_2 \,z^3$.

\section{Relation to previous results - discussion}

As a matter of fact, all known older results are based on
equations which are readily recovered, as special cases, from Eqs.
(\ref{discrete-eqmotion-x}) and (\ref{discrete-eqmotion-z}) and/or
their continuum counterparts (\ref{eqmotion-gen-continuum-x}) and
(\ref{eqmotion-gen-continuum-z}). In particular, the coupled Eqs.
(1) and (2) in Ref. \cite{Ivlev2003} are exactly recovered from
(\ref{discrete-eqmotion-x}) and (\ref{discrete-eqmotion-z}), upon
neglecting $a_{30}$, $a_{12}$, $K_1$ and $K_2$ and then evaluating
all coefficients for a Debye--type potential.

Upon switching off the coupling (i.e. setting $w \rightarrow 0$),
Eq. (\ref{eqmotion-gen-continuum-x}) above recovers exactly the
nonlinear Eq. (13) in Ref. \cite{IKPKSEPJD}, which was therein
shown to model (nonlinear) \emph{longitudinal} dust grain motion
in terms of (either Korteweg-de Vries-- \cite{Melandso, Stenflo}
or Boussinesq--type) solitons; also see Eq. (2) in
\cite{IKPKSPoPLDLWMI} (treating the formation of asymmetric
envelope modulated LDLWs) and Eq. (2) in \cite{Avinash} (keep only
first-neighbor interactions therein, to compare). In a similar
manner, considering purely \emph{transverse} motion (i.e. setting
$u \rightarrow 0$) Eqs. (\ref{discrete-eqmotion-z}) and
(\ref{eqmotion-gen-continuum-z}) herein recover exactly the
nonlinear Eqs. (7) and (8) in Ref. \cite{IKPKSPoPlast}, where they
were shown to model the amplitude modulation of TDLWs which is due
to the sheath nonlinearity. Finally, needless to say, the linear
limit recovers exactly the known equations of motion for either
purely longitudinal or purely transverse motion (i.e. considering
$a_{ij} = K_j = 0, \,\, \forall \,i, j$).

An exact treatment of the coupled evolution Eqs.
(\ref{discrete-eqmotion-x}, \ref{discrete-eqmotion-z}) -- or, at
least, the continuum system (\ref{eqmotion-gen-continuum-x},
\ref{eqmotion-gen-continuum-z}) -- seems quite a complex task to
accomplish. Even though Eq. (\ref{eqmotion-gen-continuum-x}) may
straightforward be seen as a Boussinesq--type equation
\cite{IKPKSEPJD}, which is now modified by the coupling, its
transverse counterpart (\ref{eqmotion-gen-continuum-z}) (for $u
\rightarrow 0$, say) substantially differs from any known
nonlinear model equation, bearing known exact solutions.
Therefore, we shall limit ourselves to reporting this system of
evolution equations, for the first time, thus keeping a more
thorough investigation (analytical and/or numerical) of their
fully nonlinear regime for a later report.

\section{Coupled-mode modulated wave packets}

In order to gain some insight regarding the influence of the
mode--coupling on the nonlinear profile of the dust lattice waves,
we may consider the effects which come into play when the
amplitude of the LDLWs and the TDLWs -- which are initially
uncoupled in the small amplitude (linear) limit -- is increased to
a slightly finite (i.e. non negligible) value, thus allowing for a
weak coupling between the two modes and a tractable appearance of
the signature of the (weak) nonlinearity in the dynamics.

The standard way for such an approach is via the introduction of
multiple space and time scales, viz. $X_0, X_1, X_2, ...$ and
$T_0, T_1, T_2, ...$, where $X_n = \epsilon^n x$ and $T_n =
\epsilon^n t$ ($\epsilon \ll 1$ is a smallness parameter). The
solutions are expanded as: $u = \epsilon u_1 + \epsilon^2 u_2 +
...$ (plus an analogous expression for $w$). The technical details
of the calculation are described e.g. in \cite{IKPKSPoPLDLWMI} and
will be omitted here. We shall apply this reductive perturbation
technique to the system obtained from Eqs.
(\ref{eqmotion-gen-continuum-x}, \ref{eqmotion-gen-continuum-z})
by keeping only the lowest-order nonlinear terms (i.e. omitting
the last line in both equations); we set $p_0 = 2 a_{20} r_0^3$
and $h_0 = 2 a_{02} r_0^3$ for simplicity. Note the inevitable
(and qualitatively expected) complication of the calculation due
to the different dispersion laws in the two modes \cite{inprep}.

The first-order ($\sim \epsilon$) equations are uncoupled and may
be solved by assuming $\{ u_1, w_1 \} = \{ \psi_L^{(0)},
\psi_T^{(0)} \} $ $+ [\{ \psi_L, \psi_T \} \exp i (k x - \omega t)
+ {\rm{c.c.}}]$ (complex conjugate). Upon substitution, we obtain
$\psi_T^{(0)}=0$; the remaining (3) amplitudes are left arbitrary.
This readily yields the known dispersion relations
\begin{equation}
\omega_L^2 + i \nu \omega_L = c_L^2 k^2 \biggl( 1 - \frac{k^2
r_0^2}{12} \biggr) \, , \qquad \omega_T^2 + i \nu \omega_T =
\omega_g^2 - c_T^2 k^2 \biggl( 1 - \frac{k^2 r_0^2}{12} \biggr) \,
, \label{dispersion}
\end{equation}
for the (acoustic) LDL and the (optical-like) TDL mode
respectively.

The 2nd-order ($\sim \epsilon^2$) equations contain secular
(1st-harmonic forcing) terms, whose elimination imposes a pair of
conditions in the form: $\partial \Psi_j/\partial T_1 + v_{g, j}\,
\partial \Psi_j/\partial X_1 = 0$ (where $j \in \{ 1, 2 \} \equiv
\{ L, T \}$ in the following), implying that the amplitudes
$\Psi_j$ travel at the (different) group velocities $v_{g, j}
\equiv
\partial{\omega_j}(k)/\partial k$. See that $v_{g, T} = {\omega_T}'(k)
< 0$ (the TDLW is a \emph{backward wave}), as immediately obtained
from (\ref{dispersion}b). The remaining system is then solved for
the $0$th and the $2$nd harmonic amplitudes (in $\epsilon^2$)
\cite{inprep}; the solution finally obtained is of the form:
\begin{eqnarray}
\delta x_n(t) & \approx & u(x, t) \approx \epsilon [\psi_0 +
\psi_1 \exp i (k x - \omega_1 t) + {\rm{c.c.}}] + \epsilon^2 \,
u_2^{(2)} \exp 2 i (k x - \omega_1 t) + {\rm{c.c.}}] + \mathcal
O(\epsilon^3) \nonumber \\
\delta z_n(t) &\approx & w(x, t) \approx \epsilon [\psi_2 \exp i
(k x - \omega_2 t) + {\rm{c.c.}}] + \epsilon^2 \, \{ w_2^{(0)} +
[w_2^{(2)} \exp 2 i (k x - \omega_2 t) + {\rm{c.c.}}] \} +
\mathcal O(\epsilon^3) \, .
\end{eqnarray}
We henceforth denote the significant amplitudes $u_1^{(0)}$,
$u_1^{(1)}$ and $w_1^{(1)}$ by $\Psi_0$, $\Psi_1$ and $\Psi_2$
respectively. The 2nd order correction amplitudes are
\begin{equation}
u_2^{(2)} = i k^3 \, \frac{p_0 \Psi_1^2 - h_0 \Psi_2^2}{D_2^{(L)}}
\, , \qquad w_2^{(0)} = - \frac{2 K_1}{\omega_g^2} |\Psi_1|^2 \, ,
\qquad w_2^{(2)} = - \frac{1}{D_2^{(T)}} \biggl( K_1 \Psi_2^2 + 2
i h_0 k^3 \Psi_1 \Psi_2 \biggr) \, , \label{n2l2}
\end{equation}
where $D_2^{(L)} = - c_L^2 r_0^2 k^4 + 2 i \nu \omega_L$ and
$D_2^{(T)} = - 3 \omega_g^2 + c_T^2 r_0^2 k^4 + 2 i \nu \omega_T
$. The contributions $u_2^{(1)}$, $w_2^{(1)}$ and $u_2^{(0)}$ are
left arbitrary by the algebra and were thus set to zero.

Proceeding to the 3rd-order ($\sim \epsilon^3$) equations, the
elimination of the secular terms together with zeroth order
equations provide three explicit conditions, for $\Psi_{0, 1, 2}$.
After some tedious algebra, these take the form
\begin{eqnarray}
i \biggl( \frac{\partial \Psi_1}{\partial T_2} + v_{g,
1}\frac{\partial \Psi_1}{\partial X_2} \biggr) \, + \, P_1 \,
\frac{\partial^2 \Psi_1}{\partial X_1^2} \, + \,Q_{11} \,
|\Psi_1|^2 \Psi_1 \,+ \,Q_{12} \, |\Psi_2|^2 \Psi_1 \, + (Q_{0, 1}
\Psi_1 + Q_{0, 2} \Psi_2) \frac{\partial \Psi_0}{\partial X_1} +
H_1 = 0 \nonumber \\
i \biggl( \frac{\partial \Psi_2}{\partial T_2} + v_{g,
2}\frac{\partial \Psi_2}{\partial X_2} \biggr) \, + \, P_2 \,
\frac{\partial^2 \Psi_2}{\partial X_1^2} \, + \,Q_{22} \,
|\Psi_2|^2 \Psi_2 \,+ \,Q_{21} \, |\Psi_1|^2 \Psi_2 \, +
H_2 = 0 \qquad \qquad \qquad \qquad \qquad \quad \nonumber \\
(v_{g, 1}^2 - c_L^2) \frac{\partial \Psi_0}{\partial X_1} = - p_0
k^2 |\Psi_1|^2 + h_0 k^2 |\Psi_2|^2 + C \, \, , \qquad \qquad
\qquad \qquad \qquad \qquad \label{NLSE012}
\end{eqnarray}
where $C$ is an integration constant (to be determined by the
boundary conditions). The linear dispersion terms $P_j$ are
related to the (curvature of) the dispersion relations
(\ref{dispersion}) as $P_j = {\omega_j}''(k)$ ($j = 1, 2$); the
group velocities $v_{g, j}$ were defined above \cite{comment1}.
The nonlinearity coefficients $Q_{ij}$ ($i = 0, 1, 2$, $j = 1, 2$)
and the `peculiar' contributions $H_j$ (involving cross-terms in
$\Psi_i^2 \Psi_j^*$) are too lengthy to report here \cite{inprep}.
Observe that, once $C$ is determined, one may cast Eqs.
(\ref{NLSE012}) into the form of a (modified, asymmetric) system
of coupled nonlinear Schr\"odinger equations (CNLSE). Note that we
have avoided the usual envelope (Galilean) transformation $\{ x, t
\} \rightarrow \{ x - v_{g, j} t, t \}$, since it does not
simplify this (asymmetric, with respect to $1 \leftrightarrow 2$)
system. Finally, let us point out, for rigor, that the results in
\cite{IKPKSPoPLDLWMI} and \cite{IKDPC} are exactly recovered, from
both (\ref{n2l2}) and (\ref{NLSE012}), in the appropriate --
uncoupled mode -- limits (namely, $\Psi_2 \rightarrow 0$ and
$\Psi_1 \rightarrow 0$, respectively, for LDLWs and TDLWs).

Despite the obvious analytical complication, the physical
mechanism underlying the above results is rather transparent.
There is an energy pumping effect between the zeroth-harmonic
longitudinal (displacement) mode $\Psi_0$, first put forward in
\cite{IKPKSPoPLDLWMI} (for LDLWs, yet long known in solid state
physics \cite{Tsurui}) and the modulated (low-frequency) LDL and
(high-frequency) TDL mode(s). Note the strong misfit (asymmetry)
between the dispersion laws dominating the coupled modes, despite
which -- regretfully -- no simplifying assumption may be
analytically carried out in this continuum model.

\section{Conclusion}

We have put forward a comprehensive
nonlinear model for coupled longitudinal-to-transverse
displacements in a horizontal dust mono-layer, levitated in a
sheath under the influence of gravity and an electric and/or
magnetic field. All of the above results are generic, i.e. valid
for any assumed form of the inter-grain interaction potential
$U(r)$ and the sheath potential $\Phi$, and will hopefully
contribute to the elucidation of the grain oscillatory dynamics in
dust crystals.

\section*{Appendix: Form of the coefficients for the Debye interaction
potential}

Consider the Debye potential (energy) $U_D(r) = q \phi_D(r) =
q^2\, e^{-r/\lambda_D}/r$. Defining the (positive real) lattice
parameter $\kappa = r_0/\lambda_D$, one straightforward has
\[
U'_D(r_0) = - \frac{q^2}{\lambda_D^2} \, e^{-\kappa}\, \frac{1 +
\kappa}{\kappa^2} \, , \qquad U''_D(r_0) = + \frac{2
q^2}{\lambda_D^3}\, e^{-\kappa}\, \frac{1 + \kappa+
\frac{\kappa^2}{2}}{\kappa^3} \, ,
\]
\[
U'''_D(r_0) = - \frac{6 q^2}{\lambda_D^4} \, e^{-\kappa}\, \frac{1
+ \kappa+ \frac{\kappa^2}{2} + \frac{ \kappa^3}{6}}{\kappa^4} \, ,
\qquad U''''_D(r_0) = + \frac{24 q^2}{\lambda_D^5} \,
e^{-\kappa}\, \frac{1 + \kappa + \frac{\kappa^2}{2}+
\frac{\kappa^3}{6} + \frac{ \kappa^4}{24}}{\kappa^5} \, ,
\]
where the prime denotes differentiation and $l = 1, 2, 3, ...$ is
a positive integer. Now, combining with definitions
(\ref{frequencies}, \ref{coefficients}), we have:
\[
\omega_{L, 0}^2 = \frac{2 q^2}{M \lambda_D^3} \, e^{-\kappa}\,
\frac{1 + \kappa + \kappa^2/2}{\kappa^3} \equiv c_L^2/(\kappa^2
\lambda_D^2) \,\, , \qquad \omega_{T, 0}^2 = \frac{q^2}{M
\lambda_D^3} \, e^{-\kappa}\, \frac{1 + \kappa}{\kappa^3} \equiv
c_T^2/(\kappa^2 \lambda_D^2) \,\, ,
\]
\[
p_0 \equiv 2 a_{20} \kappa^3 \lambda_D^3 = \frac{6 q^2}{M
\lambda_D} \, e^{-\kappa}\, \biggl( \frac{1}{\kappa} + 1 +
\frac{\kappa}{2}+ \frac{\kappa^2}{6} \biggr) \,\, , \qquad h_0
\equiv 2 a_{02} \kappa^3 \lambda_D^3 = \frac{3 q^2}{M \lambda_D}
\, e^{-\kappa}\, \biggl( \frac{1}{\kappa} + 1 + \frac{\kappa}{3}
\, , \biggr)\]
\[
a_{30} = \frac{q^2}{6 M \lambda_D^5} \, e^{-\kappa}\,
\frac{1}{\kappa^5} \biggl( \kappa^4 + 4 \kappa^3 + 12 \kappa^2
+ 24 \kappa + 24 \biggr) \,\, , \qquad a_{12} = \frac{q^2}{2 M
\lambda_D^5} \, e^{-\kappa}\, \frac{1}{\kappa^5} \biggl( \kappa^3
+ 5 \kappa^2 + 12 \kappa + 12 \biggr) \, \, .\] Of course,
all known previous definitions of (some of) these coefficients
(for nearest neighbour interactions; see in the references cited
in the text) are exactly recovered. Note, finally, that $\kappa$
is of the order of (or slightly higher than) unity in experiments;
therefore, all coefficients turn out to be of similar order of
magnitude, as one may check numerically.

\begin{acknowledgments}
This work was supported by the {\it{SFB591
(Sonderforschungsbereich) -- Universelles Verhalten
gleichgewichtsferner Plasmen: Heizung, Transport und
Strukturbildung}} German government Programme. Support by the
European Commission (Brussels) through the Human Potential
Research and Training Network via the project entitled: ``Complex
Plasmas: The Science of Laboratory Colloidal Plasmas and
Mesospheric Charged Aerosols'' (Contract No. HPRN-CT-2000-00140)
is also acknowledged.
\end{acknowledgments}


\begin{thebibliography}{10}
\bibitem{comment1} See that, once damping is taken into account
(via $\nu \neq 0$), all dispersion-related quantities -- namely
$\omega_j$, $v_{g, j}$ and $P_j$ ($j = 1, 2 \equiv L, T$) here --
bear an imaginary part. All of the corresponding expressions --
readily obtained from Eq. (\ref{dispersion}) and thus omitted here
-- were exactly recovered by the (tedious) algebra within our
perturbative scheme; recall, in passing, that $\Psi_{0, 1, 2}$ are
complex numbers, by definition.
\end{thebibliography}
\end{document}